\documentclass[proof]{WileyASNA-v1}

\usepackage{lipsum}

\articletype{Article Type}%

\received{}
\revised{}
\accepted{}

\begin{document}

\title{Hybrid neutron stars in the mass-radius diagram}

\author[1]{Mateusz Cierniak*}
\author[1,2,3]{David Blaschke**}

\authormark{Mateusz Cierniak \& David Blaschke}

\address[1]{\orgdiv{Institute of Theoretical Physics}, \orgname{University of Wroclaw}, \orgaddress{50-204 Wroclaw, \country{Poland}}}

\address[2]{\orgdiv{Bogoliubov Laboratory of Theoretical Physics, JINR Dubna, 141980 Dubna, Russia}}  

\address[3]{\orgdiv{National Research Nuclear University (MEPhI), 115409 Moscow, Russia}}  

\corres{* \email{mateusz.cierniak@uwr.edu.pl} \\ ** \email{david.blaschke@uwr.edu.pl}}

\abstract{
We present a systematic investigation of the possible locations for the special point (SP), a unique feature of hybrid neutron stars in the mass-radius.
The study is performed within the two-phase approach where the high-density (quark matter) phase is described by the constant-sound-speed (CSS) equation of state (EoS) and the nuclear matter phase around saturation density is varied from very soft (APR) 
to stiff (DD2 with excluded nucleon volume). 
Different construction schemes for the deconfinement transition are applied: Maxwell construction, mixed phase construction and parabolic interpolation.
We demonstrate for the first time that the SP is invariant not only 
against changing the nuclear matter EoS, but also against variation of the construction schemes for the phase transition.
Since the SP serves as a proxy for the maximum mass and accessible radii of massive hybrid stars, we draw conclusions for the limiting masses and radii of hybrid neutron stars.
}


\maketitle

\section{Introduction}

Recently, significant progress has been made in the field of nuclear astrophysics due to the observation of gravitational radiation from the inspiral phase of two colliding neutron stars, an event labeled GW170817 \citep{TheLIGOScientific:2017qsa} which was followed by observations of the kilonova event in all wavelength bands of the electromagnetic spectrum \citep{GBM:2017lvd} and marked the begin of the era of multi-messenger astronomy.  
The significance of this observation comes from the fact that, for the first time, it was possible to extract limits on the tidal deformability of the merging neutron stars and thus to provide a new constraint on the properties of dense neutron star matter like its stiffness \citep{Hinderer:2009ca}.
Further analysis of this signal resulted in limits on the radii of the constituent stars \citep{Annala:2017llu, Bauswein:2017vtn} and indicated a possible path towards detecting a phase transition inside neutron star cores from future collision events \citep{Bauswein:2018bma,Bauswein:2020aag}).

One of the most intriguing questions in modern astrophysics is whether there is a phase transition inside the core of massive neutron stars.
The answer remains elusive due to the lack of clear observational evidence, as well as the possible ambiguity in the meaning of such a signal, i.e. what did the neutron star core transition into, a cold dense quark--gluon plasma or an entirely different exotic state (cf. \cite{Marczenko:2020wlc} and references therein).
The character of the transition (i.e. first order, second order, smooth crossover etc.) also remains unknown.
Both of these carry non--trivial implications on the predicted neutron star properties.
Unfortunately, any clear answers coming from first--principle QCD calculations remain inaccessible.

An answer may come thanks to another recent development.
The NICER mission, an x--ray telescope on board of the International Space Station, was able to gather luminosity data from two pulsars, the highest mass neutron star ever observed - PSR J0740+6620 (\cite{Cromartie:2019kug, Fonseca:2021wxt, Riley:2021pdl, Miller:2021qha}) and an intermediate mass object, PSR J0030+0451 (\cite{Miller:2019cac, Riley:2019yda}).
This allowed for the derivation of limits on the star's radii. 
The measurements only marginally agree with the prediction based on the GW170817 tidal deformability limits (\cite{Abbott:2018exr}), thus resulting in a rather narrow region of probable radii for intermediate mass neutron stars \citep{Capano:2019eae}.
Combined with the novel high mass pulsar measurement, we can notice a peculiarity in the neutron star mass--radius diagram, an observation of a rather compact (soft core) set of intermediate mass stars neighbouring a less compact (stiffer, less compressible core) high mass family.

Such an observation is difficult to explain, using single phase hadronic models, while at the same time consistently describing known microscopic features of matter, see for example the hyperon puzzle (\cite{Chatterjee:2015pua}), or other important features (\cite{Yamamoto:2015lwa, Yamamoto:2017wre}), which tend to soften the high density part of the model's equation of state (EoS).
The conventional wisdom on multi--phase models is also at odds with these measurements, as typically a Maxwell construction of a first-order phase transition leads to a softening of the EoS.
Known exceptions from this paradigm are the 
interpolation construction of a crossover transition between a soft hadronic and a stiff quark matter EoS \citep{Masuda:2012ed,Baym:2017whm,Ayriyan:2021prr} and the 
two-family scenario of \cite{Drago:2020gqn} 
which is based on a nonequilibrium concept for the transition from a family of neutron stars with soft hadronic matter EoS to the family of stiff and massive strange quark star EoS.

In this manuscript, we will analyze the possibility of a phenomenological constant--speed--of--sound (CSS) model to consistently explain all of the current multi-messenger observations.
For that purpose we will employ the "Special Point" interface, a heuristic tool discovered in \cite{Yudin:2014mla} and recently studied in \cite{Cierniak:2020eyh} and \cite{Blaschke:2020vuy}.
This tool makes use of a property that is unique to two--phase models, namely the existence of a stationary point for hybrid star sequences in the mass--radius plot with respect to variations of the onset density of the second phase. It has been demonstrated in \cite{Cierniak:2020eyh}, that this point is insensitive to the other phase chosen. Additionally, a precise relation between this point's mass and the maximum mass was derived, thus demonstrating the capability of this feature in estimating model parameters based on observational constraints.

The manuscript is organized as follows. In Section \ref{sec::2}, subsection \ref{sec::2::sub::1} will focus on the phenomenological quark matter equation of state. Subsection \ref{sec::2::sub::2} will be devoted to the phase transition construction linking the separate hadronic and quark equations of state. Section \ref{sec::3} will discuss the systematics of the Special Point feature. Section \ref{sec::4} will conclude with a summary of findings and discussion.

%%%%%%%%%%%%%%%%%%%%%%%%%%%%%%%%%%%%%%%%%%%%%%%%%%%%%%%%% 

\section{Equation of state and phase transition construction}
\label{sec::2}

We will consider hybrid equations of state (EoS) within the two-phase model scheme, where hadronic and quark matter EoS are modeled separately and the resulting hybrid EoS is obtained by a phase transition construction. 

\subsection{Constant sound speed EoS}
\label{sec::2::sub::1}

For the deconfined quark phase, a class of constant speed of sound (CSS) quark matter models will be used. This equation of state in the form of pressure as a function of the baryochemical potential reads (see also the Appendix of \citep{Alford:2013aca}), 
\begin{equation} 
\label{css3}
    P(\mu)=A(\mu/\mu_0)^{1+c_s^{-2}}-B,
\end{equation}
where the model parameters $A$, $B$ and $c_s^2$ are constant and $\mu_0=1$ GeV.
The baryon density $n$ follows from the canonical relation
\begin{equation}
\label{eq:density}
    \frac{d P(\mu)}{d \mu}= n(\mu)
    =(1+c_s^{-2})\frac{A}{\mu_0}\left(\frac{\mu}{\mu_0}\right)^{c_s^{-2}}.
\end{equation}
Using the above, we arrive at the energy density
\begin{equation}
    \varepsilon=\mu n-P=B + c_s^{-2} A(\mu/\mu_0)^{1+c_s^{-2}}.
\end{equation}
The relation between pressure and energy density takes the form
\begin{equation} \label{css6}
   P = c_s^2  \varepsilon - (1+c_s^2)B.
\end{equation}
%From Eq.~\ref{css6} we see, that $\varepsilon_0=(1+c_s^{-2})B$. 
The pressure slope parameter $A$ does not affect the relation between pressure and energy density, but its values should be limited to a range that produces a non-negative density jump at the phase transition.
It has been shown by \cite{Zdunik:2012dj} that the above CSS model fits well the EoS of color superconducting quark matter in both, the 2SC and the CFL phases which were obtained from a selfconsistent solution of the three-flavor NJL model in the mean field approximation \citep{Blaschke:2005uj,Klahn:2013kga}.

\subsection{Phase transition constructions}
\label{sec::2::sub::2}
 
In this work, we use the replacement interpolation method (RIM) described in \cite{Abgaryan:2018gqp} in which a simple modification of the Maxwell construction is employed. 
Since the EoS of hadronic and quark matter phases are described with the relation between the pressure and chemical potential, $P_H(\mu)$ and $P_Q(\mu)$ respectively, the effective mixed phase EoS $P_M(\mu)$ can be described simply by an interpolated function between these two phases. 
It requires that the interpolated pressure coincides with the values of hadronic and quark matter pressures at the lower and upper borders of the mixed phase where not only the continuity of pressures but also the thermodynamic constraint of positive slope of density versus chemical potential, i.e., ${\partial n_M}/{\partial \mu_M} = {\partial^2 P_M}/{\partial \mu_M^2}  > 0$ shall be fulfilled.
Moreover, we require the causality condition to be fulfilled, namely that the adiabatic speed of sound at zero frequency, 
$c_s = \sqrt{{\partial P}/{\partial \epsilon}}$, does not exceed the speed of light. A simple and reasonable function to interpolate the pressure is a polynomial function which smoothly joins the pressure curves of two phases.
This method has been developed in \cite{Ayriyan:2017tvl} and applied to the question of robustness of NS mass twins against mixed phase effects in 
\cite{Ayriyan:2017nby}. We repeat here the basic steps of its derivation following \cite{Ayriyan:2017tvl} and \cite{Ayriyan:2017nby}.

\begin{figure}[!htb]
    \includegraphics[scale=0.5]{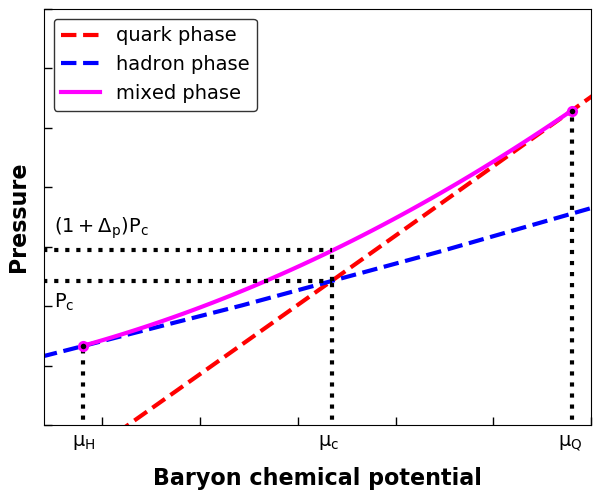}
\caption{
Hybrid EoS construction (replacement interpolation method (RIM) \cite{Abgaryan:2018gqp,Ayriyan:2021prr}) when the hadronic and quark matter functions for $P(\mu)$ would cross in the correct way, i.e. quark matter having a smaller pressure than hadronic matter at low chemical potentials and vice-versa at high chemical potentials.}
\label{fig:maxwell}
\end{figure}

The value of the critical baryochemical potential $\mu_c$ for which the phases are in mechanical and chemical equilibrium with each other is obtained from the Gibbs condition of phase equilibrium
  \begin{equation}
  P_Q(\mu_c)~=~P_H(\mu_c)~=~P_c~.
   \label{eq:mechequi}
  \end{equation}
For the pressure of the mixed phase a parabolic ansatz is considered
here
\begin{equation}
 P_M(\mu)~=~\alpha_2(\mu-\mu_c)^2~+~\alpha_1(\mu-\mu_c)~+~(1+\Delta_P)P_c,
  \label{eq:pppressure}
\end{equation}
where $\Delta_P$ is a free parameter which determines the pressure of mixed phase at $\mu_c$
\begin{equation}
 P_M(\mu_c)~=~P_c+\Delta_P~=~P_M~,~\Delta_P=\Delta P/P_c,
  \label{eq:addpressure}
\end{equation}
see Fig.~\ref{fig:maxwell}
The value of $\Delta_P$ is related to the surface tension between two phases such that the vanishing $\Delta_P$ corresponds to a minimal value of the surface tension for which the transition becomes equivalent to that of the Maxwell construction.
The quantitative relation between $\Delta_P$ and the surface tension has been given in \cite{Maslov:2018ghi}
for a selection of hybrid EoS cases and it shows that a value of $\Delta_P \approx 0.05-0.07$ corresponds to a vanishing surface tension and thus a construction defined in \cite{Glendenning:1992vb}. 

According to the Gibbs conditions for phase equilibrium (\ref{eq:mechequi}) 
the pressures and their derivative of order $k$ have to satisfy the continuity conditions
\begin{eqnarray}
P_H(\mu_{H}) &=&  P_M(\mu_{H})~,\\
P_Q(\mu_{Q}) &=&  P_M(\mu_{Q})~,\\
\frac{\partial^k}{\partial\mu^k}P_H(\mu_{H}) &=&  \frac{\partial^k}{\partial\mu^k}P_M(\mu_{H})~,\\
\frac{\partial^k}{\partial\mu^k}P_Q(\mu_{Q}) &=&  \frac{\partial^k}{\partial\mu^k}P_M(\mu_{Q})~.
\label{eq:continuity}
\end{eqnarray}
For the ansatz of Eq.~(\ref{eq:pppressure}), $k=1$. These continuity conditions allow us to find analytical expressions for $\alpha_1$ and $\alpha_2$,
\begin{equation}
    \alpha_1
    =
    \frac
    {-2\kappa_1+\kappa_2(\mu_c-\mu_H)}
    {2(\mu_c-\mu_Q)(\mu_H-\mu_Q)},
\end{equation}
\begin{equation}
    \alpha_2
    =
    \frac
    {-2\kappa_1+\kappa_2(\mu_c-\mu_Q)}
    {2(\mu_c-\mu_H)(\mu_H-\mu_Q)},
\end{equation}
with
\begin{equation}
    \begin{cases}
    \kappa_1=n_Q(\mu_c-\mu_Q)-n_H(\mu_c-\mu_H)+P_Q-P_H,\\
    \kappa_2=n_Q-n_H,
    \end{cases}
\end{equation}
which can be used to numerically derive $\mu_{H}$ and $\mu_{Q}$.

\begin{figure}[!htb]
    \includegraphics[scale=0.5]{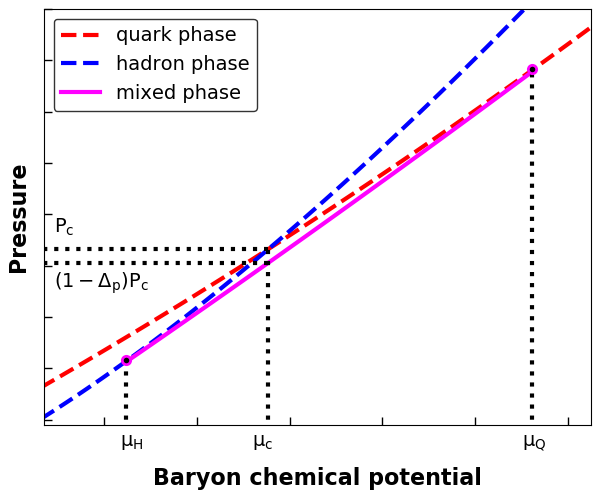}
\caption{
Hybrid EoS construction (replacement interpolation method (RIM) \citep{Abgaryan:2018gqp,Ayriyan:2021prr}) when the hadronic and quark matter functions for $P(\mu)$ would cross in the wrong way, i.e. quark matter having a larger pressure than hadronic matter at low chemical potentials and vice-versa at high chemical potentials.}
\label{fig:rim}
\end{figure}

It is worth mentioning that this interpolation method can be applied not only to the usual phase transition from the hadronic phase to the quark phase with 
$\Delta_P > 0$ but also to the case where applying the principle of maximum pressure to the crossing of the pressure vs. chemical potential curves for quark and hadron matter would correspond to a nonphysical reconfinement transition \citep{Zdunik:2012dj} from quark phase to hadronic phase with $\Delta_P < 0$, see Fig.~\ref{fig:rim}. Such a situation  corresponds to the case where an interpolation has been suggested to describe a crossover transition \citep{Masuda:2012ed,Baym:2017whm,Ayriyan:2021prr}.
Such a construction was explored in \cite{Shahrbaf:2020uau} and will be analyzed in the context of the mass--radius special point in the following section.

\section{Location of the special point and its properties}
\label{sec::3}

\begin{figure}[!htb]
    \includegraphics[scale=0.5]{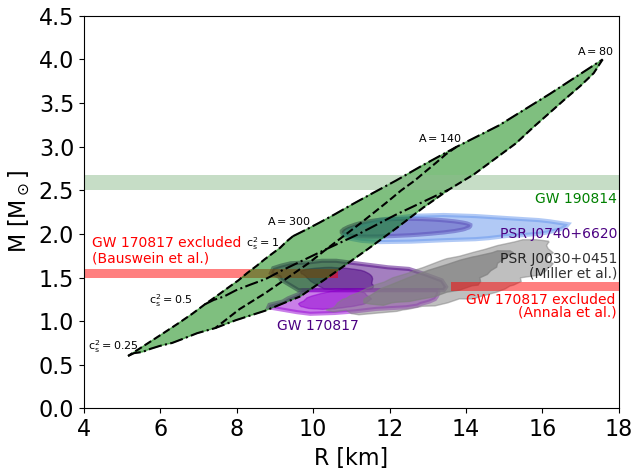}
\caption{
The $M-R$ diagram with the grid of special point positions obtained by varying the values of the CSS quark matter model parameters  $A$ and $c_s^2$.
For a comparison we show the new 1-$\sigma$ mass-radius constraints from the NICER analysis of observations of the massive pulsar PSR J0740+6620 \citep{Fonseca:2021wxt} as dark blue \citep{Riley:2021pdl} and light blue \cite{Miller:2021qha} regions. 
Additionally, we mark by red bars the excluded regions for a lower limit \citep{Bauswein:2017vtn} and an upper limit \citep{Annala:2017llu} on the radius deduced from the gravitational wave observation GW170817.
The grey regions are the 1-$\sigma$ and 2-$\sigma$ areas from the NICER mass-radius measurement on PSR J0030+0451 \citep{Miller:2019cac} and the light green band is the mass range of the lighter object in the binary merger GW190814 \citep{Abbott:2020khf}.
}
\label{fig:grid}
\end{figure}

\begin{figure}[!htb]
    \includegraphics[scale=0.5]{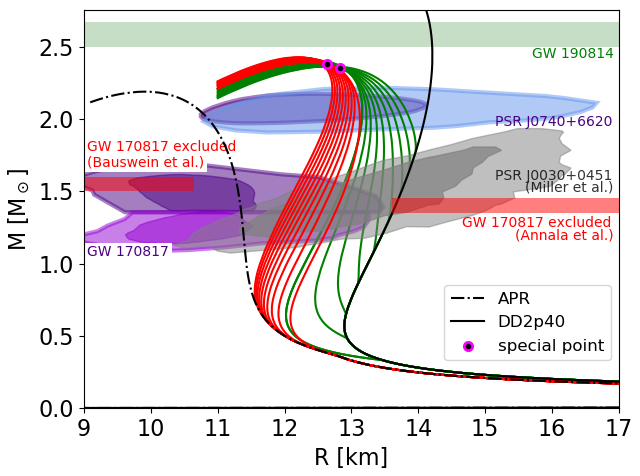}
\caption{The $M-R$ sequences for hybrid star EoS obtained by a Maxwell construction for a DD2p40 EoS (black solid line) with the CSS model (green solid lines), as well as the RIM mixed phase construction from the softer APR EoS (black dot--dashed line) to the same CSS model (red solid lines) with $A=81$ MeV/fm$^3$, $c_s^2=0.46$, and $65 < B~ [{\rm MeV/fm}^3] < 85$. The parameter of the RIM construction is set to $\Delta_p=-5\%$.  }
\label{fig:massradius_apr}
\end{figure}

In \cite{Cierniak:2020eyh} the region in the M--R diagram has been explored where the SP point (\cite{Yudin:2014mla}) can be located, while appearing on the stable part of the hybrid neutron star branch.
We further improve the systematics of this study by taking into account a lower bound on the A parameter, corresponding to a RIM mixed phase for which the onset density is equal to the nuclear saturation density $n_0=0.16~$fm$^{-3}$.
From this condition, together with the chemical potential at the saturation density $\mu(n_0)=\xi \mu_0$, follows from Eq.~(\ref{eq:density}) for the limiting case $c_s^2=1$
the minimal value of $A$,
\begin{equation}
    A_{\rm min}=\frac{n_0 \mu_0}{2\xi} = 81~{\rm MeV/fm}^3~,
\end{equation}
when for typical nuclear EoS $\xi=0.98$.
By combining this limit with the relation between the mass $M_{\rm SP}$ of the SP and $M_{\rm max}$, that has been found by \cite{Blaschke:2020vuy}, we can estimate the minimum value of the maximum mass ($M_{\rm max}=M_{\rm SP}$) for a given choice of the speed of sound parameter.
The extent of this estimate is visible in Fig.~\ref{fig:grid}, with the resulting prediction on $M_{\rm max}$ and the range of radii to be expected at the mass of PSR J0740+6620 listed in Table~\ref{tab:1}.
The minimal radius $R_{J0740, \rm min}$ is attained for the case when 
$M_{\rm SP}=M_{J0740}=2.08\pm 0.07~M_\odot$ \citep{Fonseca:2021wxt}, while the maximum radius $R_{J0740, \rm max}$ is estimated from the stiffest EoS parametrization that does not violate the upper radius limit of \citep{Annala:2017llu}.
%from the crossing of $M(R)$ sequences with the minimal onset density $n_{\rm onset}=n_0$ for the extreme position of $M_{SP}(c_s^2,A)=M_{SP}(c_s^2,A_{\rm min})$, for both cases of the transition construction. Namely, for the soft hadronic EoS (APR) constructed with the RIM and negative $\Delta_p$ parameter and for the stiff hadronic EoS DD2p40 \citep{Typel:2016srf,Alvarez-Castillo:2016oln} with the RIM and positive $\Delta_p$ parameter, for the earliest possible onset $n_{\rm onset}=n_0$.
For an illustration, see Fig.~\ref{fig:massradius_apr}.
\begin{table}
\begin{center}
\begin{tabular}{|c|c|c|c|}
\hline
$c_s^2$ & $M_{\rm SP}$ & $R_{J0740, \rm min}$&$R_{J0740, \rm max}$\\
 & [$M_\odot$] &  [km] & [km] \\
\hline
0.35 & 1.82 & - & - \\
0.40 & 2.07 & 12.18 & 12.29 \\
0.45 & 2.30 & 11.84 & 13.41 \\
0.50 & 2.50 & 11.56 & 13.91 \\
0.55 & 2.68 & 11.30 & 14.20 \\
0.60 & 2.86 & 11.05 & 14.45 \\
0.70 & 3.22 & 10.67 & 14.67 \\
1.00 & 4.00 & 9.95 & 14.84 \\
\hline
\end{tabular}
\caption{Maximum neutron star masses $M_{\rm max}$ and expected range of radii  $R_{J0740, \rm min}$, $R_{J0740, \rm max}$ at the mass of PSR J0740+6620 \citep{Fonseca:2021wxt} as derived from the CSS model with a given value of the speed--of--sound parameter $c_s^2$.
See text for details.}
\label{tab:1}
\end{center}
\end{table}

\begin{figure}[!htb]
    \includegraphics[scale=0.5]{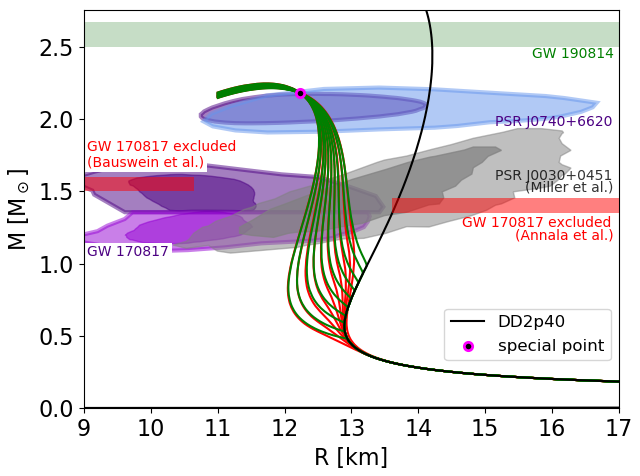}
\caption{
The $M-R$ sequences for hybrid star EoS obtained by a Maxwell construction for a DD2p40 EoS (black solid line) with the CSS model (green solid lines) for $A=91$ MeV/fm$^3$, and $c_s^2=0.46$, with different onset masses for quark deconfinement obtained by varying the bag pressure in the range $83 < B~ [{\rm MeV/fm}^3] < 87$.
All hybrid sequences go through the special point at $M_{SP}=2.17~M_\odot$ 
and $R_{SP}=12.23$ km, which remains unaffected when the Maxwell construction gets replaced by a mixed phase construction for the pressure offset $\Delta_p=4\%$ (red solid lines) that modifies the behaviour of the sequences in the vicinity of the onset of deconfinement.
}
\label{fig:massradius}
\end{figure}

As a next step, we revise the initial observation from \cite{Cierniak:2020eyh}, that the SP is insensitive to the choice of hadronic EoS, by introducing a variation of the phase transition construction. As seen in Fig.~\ref{fig:massradius}, a simple interpolated mixed phase (Fig.~\ref{fig:maxwell}) results in a meaningful change of the mass--radius relation only in the area closest to the onset mass. The SP location remains unchanged. By exploring a more radical construction from \cite{Shahrbaf:2020uau}, depicted in Fig.~\ref{fig:rim}, we arrive at a somewhat unconventional mass--radius relation (shown in Fig.~\ref{fig:massradius_apr}), motivated by the recent developments in multi--messenger astronomy, specifically the interplay between the likelihood of rather compact $1.0-1.5$ $M_\odot$ range neutron stars (\cite{Abbott:2018exr}) and the likely low--compactness of PSR J0740 (\cite{Miller:2021qha, Riley:2021pdl}). By exploring a transition from a borderline soft hadronic EoS (APR) compared to the stiff DD2p40, we immediately notice that the discrepancy in the SP mass and radius remains consistent with the findings in \cite{Cierniak:2020eyh}, thus reafirming the approximate invariance of the SP to the choice of hadronic EoS, as well as demonstrating an insensitivity to the phase transition construction.

\section{Discussion and Conclusions}
\label{sec::4}

In this work we have extended the investigations of the SP of hybrid star sequences and its properties which we started in \citep{Cierniak:2020eyh,Blaschke:2020vuy} following up on the initial study \cite{Yudin:2014mla}.
In particular, we demonstrated for the first time that the SP is also invariant when the phase transition construction within the class of two-phase approaches is altered from the traditional Maxwell construction to the replacement interpolation method using a parabolic interpolation with the lower limit on the onset of the mixed phase at $n_{\rm onset}\ge n_0$ and demanding causality ($c_s^2\le 1$) as well as mechanical stability ($dn/d\mu \ge 0$ everywhere) for the resulting hybrid EoS.  
The RIM allows for constructing the transition from a stiff hadronic 
phase to a softer quark matter phase (for non-negative mixed-phase parameter $\Delta_p\ge 0$) as well as from a soft hadronic phase to a stiffer quark matter one (for a negative interpolation parameter
$\Delta_p < 0$). 
The latter construction follows in spirit the idea of 
\citep{Masuda:2012ed}, but with a finite range of densities for the mixed phase, see also \cite{Ayriyan:2021prr}. 

This situation allows for two new types of masquerade effect
\cite{Alford:2004pf}, namely when a hybrid EoS of the first kind (stiff hadronic - soft quark matter) masquerades like a soft hadronic one (see, e.g., \cite{Blaschke:2020qqj}) and one of the second kind (soft hadronic - stiff quark matter) which masquerades like a stiff hadronic one (see, e.g., \cite{Blaschke:2020vuy}). 
Such situations can now be systematically generated and investigated with the help of the SP that should be chosen to lie on that hadronic EoS which should appear as a masquerade of a suitably chosen hybrid EoS sequence belonging to that same SP.

The second kind of masquerade belongs to a change in paradigm. 
It is not necessarily true that a hybrid EoS should be softer and thus lead to more compact hybrid star configurations than its purely hadronic counterpart.
In view of this new fact, several scenarios for identifying a deconfinement phase transition in compact star astrophysics must be revisited!
For example, the conclusion of \cite{Pang:2021jta} that a strong first-order phase transition should be disfavoured appears to be premature as it did not account for this masquerade effect with a strong and early phase transition.
Furthermore, the deconfinement signature that was predicted for the postmerger gravitational wave signal \citep{Bauswein:2018bma,Bauswein:2020aag} 
could be concerned by the new type of masquerade effect. 

In the present work, we have also provided a first systematic cartography of the special point location on the $M-R$ diagram using just two physical parameters, the squared sound speed $c_s^2$ and the quasiparticle pressure at the fiducial chemical potential $A=P(\mu_0) + B$, see Fig.~\ref{fig:grid}.
This allows to conclude that very massive objects like the lighter
companion of GW190814 with $M=2.58^{+0.08}_{-0.09}~M_\odot$ \citep{Abbott:2020khf} could well be hybrid stars. This requires only moderate stiffness of the CSS EoS with $c_s^2 \gtrsim 0.5$.
For such moderate sound speed values, the radius range expected for 
hybrid stars with the mass of PSR J0740+6620 corresponds very nicely to the NICER radius measurements \cite{Miller:2021qha,Riley:2021pdl}, while realistic models of purely hadronic EoS (including hyperons) turn out to predict too small stars above $2~M_\odot$.
We note that recently \cite{Somasundaram:2021ljr} within a first-order phase transition scenario and a CSS quark matter EoS have also obtained  very massive hybrid stars with large radii.
After completion of this work, an extensive study appeared \citep{Tan:2021ahl} which discusses the possibility of hybrid star sequences with high maximum masses and large radii obtained by either a first-order transition or a crossover construction. 
These are particular cases of the more general systematic study we presented here. 

In conclusion, with the present systematic study of the distribution and cartography of the special point in the mass-radius diagram we have proven that hybrid neutron stars can exist in the region of high masses and large radii which is inaccessible to realistic models of purely hadronic neutron stars.

%\newpage
\subsection*{Acknowledgements}
M.C. and D.B. acknowledge support from the Polish National Science Center under grant No. 2019/33/B/ST9/03059. 
D.B. acknowledges support from the Russian Foundation for Basic Research under grant No. 18-02-40137.

\bibliography{main}

\begin{thebibliography}{}

\bibitem [\protect \citeauthoryear {%
B\BPBI P.~Abbott%
\ \protect \BOthers {.}}{%
B\BPBI P.~Abbott%
\ \protect \BOthers {.}}{%
{\protect \APACyear {2017}}%
{\protect \APACexlab {{\protect \BCnt {1}}}}}]{%
TheLIGOScientific:2017qsa}
\APACinsertmetastar {%
TheLIGOScientific:2017qsa}%
\begin{APACrefauthors}%
Abbott, B\BPBI P.%
\BCBT {}\ \BOthersPeriod {.}
\end{APACrefauthors}%
\unskip\
\newblock
\APACrefYearMonthDay{2017{\protect \BCnt {1}}}{}{},
\newblock
\unskip
\newblock
\APACjournalVolNumPages{Phys. Rev. Lett.}{119}{16}{161101}.
\newblock
\begin{APACrefDOI} \doi{10.1103/PhysRevLett.119.161101} \end{APACrefDOI}
\PrintBackRefs{\CurrentBib}

\bibitem [\protect \citeauthoryear {%
B\BPBI P.~Abbott%
\ \protect \BOthers {.}}{%
B\BPBI P.~Abbott%
\ \protect \BOthers {.}}{%
{\protect \APACyear {2017}}%
{\protect \APACexlab {{\protect \BCnt {2}}}}}]{%
GBM:2017lvd}
\APACinsertmetastar {%
GBM:2017lvd}%
\begin{APACrefauthors}%
Abbott, B\BPBI P.%
\BCBT {}\ \BOthersPeriod {.}
\end{APACrefauthors}%
\unskip\
\newblock
\APACrefYearMonthDay{2017{\protect \BCnt {2}}}{}{},
\newblock
\unskip
\newblock
\APACjournalVolNumPages{Astrophys. J. Lett.}{848}{2}{L12}.
\newblock
\begin{APACrefDOI} \doi{10.3847/2041-8213/aa91c9} \end{APACrefDOI}
\PrintBackRefs{\CurrentBib}

\bibitem [\protect \citeauthoryear {%
B\BPBI P.~Abbott%
\ \protect \BOthers {.}}{%
B\BPBI P.~Abbott%
\ \protect \BOthers {.}}{%
{\protect \APACyear {2018}}%
}]{%
Abbott:2018exr}
\APACinsertmetastar {%
Abbott:2018exr}%
\begin{APACrefauthors}%
Abbott, B\BPBI P.%
\BCBT {}\ \BOthersPeriod {.}
\end{APACrefauthors}%
\unskip\
\newblock
\APACrefYearMonthDay{2018}{}{},
\newblock
\unskip
\newblock
\APACjournalVolNumPages{Phys. Rev. Lett.}{121}{16}{161101}.
\newblock
\begin{APACrefDOI} \doi{10.1103/PhysRevLett.121.161101} \end{APACrefDOI}
\PrintBackRefs{\CurrentBib}

\bibitem [\protect \citeauthoryear {%
R.~Abbott%
\ \protect \BOthers {.}}{%
R.~Abbott%
\ \protect \BOthers {.}}{%
{\protect \APACyear {2020}}%
}]{%
Abbott:2020khf}
\APACinsertmetastar {%
Abbott:2020khf}%
\begin{APACrefauthors}%
Abbott, R.%
, Abbott, T\BPBI D.%
, Abraham, S.%
\ et al.\end{APACrefauthors}%
\unskip\
\newblock
\APACrefYearMonthDay{2020}{}{},
\newblock
\unskip
\newblock
\APACjournalVolNumPages{Astrophys. J. Lett.}{896 (2)}{}{L44}.
\PrintBackRefs{\CurrentBib}

\bibitem [\protect \citeauthoryear {%
Abgaryan%
, Alvarez-Castillo%
, Ayriyan%
, Blaschke%
\BCBL {}\ \BBA {} Grigorian%
}{%
Abgaryan%
\ \protect \BOthers {.}}{%
{\protect \APACyear {2018}}%
}]{%
Abgaryan:2018gqp}
\APACinsertmetastar {%
Abgaryan:2018gqp}%
\begin{APACrefauthors}%
Abgaryan, V.%
, Alvarez-Castillo, D.%
, Ayriyan, A.%
, Blaschke, D.%
\BCBL {}\ \BBA {} Grigorian, H.%
\end{APACrefauthors}%
\unskip\
\newblock
\APACrefYearMonthDay{2018}{}{},
\newblock
\unskip
\newblock
\APACjournalVolNumPages{Universe}{4}{9}{94}.
\newblock
\begin{APACrefDOI} \doi{10.3390/universe4090094} \end{APACrefDOI}
\PrintBackRefs{\CurrentBib}

\bibitem [\protect \citeauthoryear {%
M.~Alford%
, Braby%
, Paris%
\BCBL {}\ \BBA {} Reddy%
}{%
M.~Alford%
\ \protect \BOthers {.}}{%
{\protect \APACyear {2005}}%
}]{%
Alford:2004pf}
\APACinsertmetastar {%
Alford:2004pf}%
\begin{APACrefauthors}%
Alford, M.%
, Braby, M.%
, Paris, M\BPBI W.%
\BCBL {}\ \BBA {} Reddy, S.%
\end{APACrefauthors}%
\unskip\
\newblock
\APACrefYearMonthDay{2005}{}{},
\newblock
\unskip
\newblock
\APACjournalVolNumPages{Astrophys. J.}{629}{}{969--978}.
\newblock
\begin{APACrefDOI} \doi{10.1086/430902} \end{APACrefDOI}
\PrintBackRefs{\CurrentBib}

\bibitem [\protect \citeauthoryear {%
M\BPBI G.~Alford%
, Han%
\BCBL {}\ \BBA {} Prakash%
}{%
M\BPBI G.~Alford%
\ \protect \BOthers {.}}{%
{\protect \APACyear {2013}}%
}]{%
Alford:2013aca}
\APACinsertmetastar {%
Alford:2013aca}%
\begin{APACrefauthors}%
Alford, M\BPBI G.%
, Han, S.%
\BCBL {}\ \BBA {} Prakash, M.%
\end{APACrefauthors}%
\unskip\
\newblock
\APACrefYearMonthDay{2013}{}{},
\newblock
\unskip
\newblock
\APACjournalVolNumPages{Phys. Rev. D}{88 (8)}{}{083013}.
\PrintBackRefs{\CurrentBib}

\bibitem [\protect \citeauthoryear {%
Annala%
, Gorda%
, Kurkela%
\BCBL {}\ \BBA {} Vuorinen%
}{%
Annala%
\ \protect \BOthers {.}}{%
{\protect \APACyear {2018}}%
}]{%
Annala:2017llu}
\APACinsertmetastar {%
Annala:2017llu}%
\begin{APACrefauthors}%
Annala, E.%
, Gorda, T.%
, Kurkela, A.%
\BCBL {}\ \BBA {} Vuorinen, A.%
\end{APACrefauthors}%
\unskip\
\newblock
\APACrefYearMonthDay{2018}{}{},
\newblock
\unskip
\newblock
\APACjournalVolNumPages{Phys. Rev. Lett.}{120 (17)}{}{172703}.
\PrintBackRefs{\CurrentBib}

\bibitem [\protect \citeauthoryear {%
Ayriyan%
\ \protect \BOthers {.}}{%
Ayriyan%
\ \protect \BOthers {.}}{%
{\protect \APACyear {2018}}%
}]{%
Ayriyan:2017nby}
\APACinsertmetastar {%
Ayriyan:2017nby}%
\begin{APACrefauthors}%
Ayriyan, A.%
, Bastian, N\BPBI U.%
, Blaschke, D.%
, Grigorian, H.%
, Maslov, K.%
\BCBL {}\ \BBA {} Voskresensky, D\BPBI N.%
\end{APACrefauthors}%
\unskip\
\newblock
\APACrefYearMonthDay{2018}{}{},
\newblock
\unskip
\newblock
\APACjournalVolNumPages{Phys. Rev. C}{97}{4}{045802}.
\newblock
\begin{APACrefDOI} \doi{10.1103/PhysRevC.97.045802} \end{APACrefDOI}
\PrintBackRefs{\CurrentBib}

\bibitem [\protect \citeauthoryear {%
Ayriyan%
\ \protect \BOthers {.}}{%
Ayriyan%
\ \protect \BOthers {.}}{%
{\protect \APACyear {2021}}%
}]{%
Ayriyan:2021prr}
\APACinsertmetastar {%
Ayriyan:2021prr}%
\begin{APACrefauthors}%
Ayriyan, A.%
, Blaschke, D.%
, Grunfeld, A\BPBI G.%
, Alvarez-Castillo, D.%
, Grigorian, H.%
\BCBL {}\ \BBA {} Abgaryan, V.%
\end{APACrefauthors}%
\unskip\
\newblock
\APACrefYearMonthDay{2021}{}{},
\newblock
\APACrefbtitle {{Bayesian analysis of multimessenger M-R data with interpolated
  hybrid EoS}.} {{Bayesian analysis of multimessenger M-R data with
  interpolated hybrid EoS}.}
\newblock
\APACrefnote{arXiv:2102.13485 [astro-ph.HE]}
\PrintBackRefs{\CurrentBib}

\bibitem [\protect \citeauthoryear {%
Ayriyan%
\ \BBA {} Grigorian%
}{%
Ayriyan%
\ \BBA {} Grigorian%
}{%
{\protect \APACyear {2018}}%
}]{%
Ayriyan:2017tvl}
\APACinsertmetastar {%
Ayriyan:2017tvl}%
\begin{APACrefauthors}%
Ayriyan, A.%
\BCBT {}\ \BBA {} Grigorian, H.%
\end{APACrefauthors}%
\unskip\
\newblock
\APACrefYearMonthDay{2018}{}{},
\newblock
\unskip
\newblock
\APACjournalVolNumPages{EPJ Web Conf.}{173}{}{03003}.
\newblock
\begin{APACrefDOI} \doi{10.1051/epjconf/201817303003} \end{APACrefDOI}
\PrintBackRefs{\CurrentBib}

\bibitem [\protect \citeauthoryear {%
Bauswein%
\ \protect \BOthers {.}}{%
Bauswein%
\ \protect \BOthers {.}}{%
{\protect \APACyear {2019}}%
}]{%
Bauswein:2018bma}
\APACinsertmetastar {%
Bauswein:2018bma}%
\begin{APACrefauthors}%
Bauswein, A.%
, Bastian, N\BHBI U\BPBI F.%
, Blaschke, D\BPBI B.%
, Chatziioannou, K.%
, Clark, J\BPBI A.%
, Fischer, T.%
\BCBL {}\ \BBA {} Oertel, M.%
\end{APACrefauthors}%
\unskip\
\newblock
\APACrefYearMonthDay{2019}{}{},
\newblock
\unskip
\newblock
\APACjournalVolNumPages{Phys. Rev. Lett.}{122 (6)}{}{061102}.
\PrintBackRefs{\CurrentBib}

\bibitem [\protect \citeauthoryear {%
Bauswein%
\ \protect \BOthers {.}}{%
Bauswein%
\ \protect \BOthers {.}}{%
{\protect \APACyear {2020}}%
}]{%
Bauswein:2020aag}
\APACinsertmetastar {%
Bauswein:2020aag}%
\begin{APACrefauthors}%
Bauswein, A.%
, Blacker, S.%
, Vijayan, V.%
\ et al.\end{APACrefauthors}%
\unskip\
\newblock
\APACrefYearMonthDay{2020}{}{},
\newblock
\unskip
\newblock
\APACjournalVolNumPages{Phys. Rev. Lett.}{125 (14)}{}{141103}.
\PrintBackRefs{\CurrentBib}

\bibitem [\protect \citeauthoryear {%
Bauswein%
, Just%
, Janka%
\BCBL {}\ \BBA {} Stergioulas%
}{%
Bauswein%
\ \protect \BOthers {.}}{%
{\protect \APACyear {2017}}%
}]{%
Bauswein:2017vtn}
\APACinsertmetastar {%
Bauswein:2017vtn}%
\begin{APACrefauthors}%
Bauswein, A.%
, Just, O.%
, Janka, H\BHBI T.%
\BCBL {}\ \BBA {} Stergioulas, N.%
\end{APACrefauthors}%
\unskip\
\newblock
\APACrefYearMonthDay{2017}{}{},
\newblock
\unskip
\newblock
\APACjournalVolNumPages{Astrophys. J. Lett.}{850 (2)}{}{L34}.
\PrintBackRefs{\CurrentBib}

\bibitem [\protect \citeauthoryear {%
Baym%
\ \protect \BOthers {.}}{%
Baym%
\ \protect \BOthers {.}}{%
{\protect \APACyear {2018}}%
}]{%
Baym:2017whm}
\APACinsertmetastar {%
Baym:2017whm}%
\begin{APACrefauthors}%
Baym, G.%
, Hatsuda, T.%
, Kojo, T.%
, Powell, P\BPBI D.%
, Song, Y.%
\BCBL {}\ \BBA {} Takatsuka, T.%
\end{APACrefauthors}%
\unskip\
\newblock
\APACrefYearMonthDay{2018}{}{},
\newblock
\unskip
\newblock
\APACjournalVolNumPages{Rept. Prog. Phys.}{81}{5}{056902}.
\newblock
\begin{APACrefDOI} \doi{10.1088/1361-6633/aaae14} \end{APACrefDOI}
\PrintBackRefs{\CurrentBib}

\bibitem [\protect \citeauthoryear {%
Blaschke%
, Ayriyan%
, Alvarez-Castillo%
\BCBL {}\ \BBA {} Grigorian%
}{%
Blaschke%
\ \protect \BOthers {.}}{%
{\protect \APACyear {2020}}%
}]{%
Blaschke:2020qqj}
\APACinsertmetastar {%
Blaschke:2020qqj}%
\begin{APACrefauthors}%
Blaschke, D.%
, Ayriyan, A.%
, Alvarez-Castillo, D\BPBI E.%
\BCBL {}\ \BBA {} Grigorian, H.%
\end{APACrefauthors}%
\unskip\
\newblock
\APACrefYearMonthDay{2020}{}{},
\newblock
\unskip
\newblock
\APACjournalVolNumPages{Universe}{6 (6)}{}{81}.
\PrintBackRefs{\CurrentBib}

\bibitem [\protect \citeauthoryear {%
Blaschke%
\ \BBA {} Cierniak%
}{%
Blaschke%
\ \BBA {} Cierniak%
}{%
{\protect \APACyear {2021}}%
}]{%
Blaschke:2020vuy}
\APACinsertmetastar {%
Blaschke:2020vuy}%
\begin{APACrefauthors}%
Blaschke, D.%
\BCBT {}\ \BBA {} Cierniak, M.%
\end{APACrefauthors}%
\unskip\
\newblock
\APACrefYearMonthDay{2021}{}{},
\newblock
\unskip
\newblock
\APACjournalVolNumPages{Astron. Nachr.}{342}{1-2}{227--233}.
\newblock
\begin{APACrefDOI} \doi{10.1002/asna.202113909} \end{APACrefDOI}
\PrintBackRefs{\CurrentBib}

\bibitem [\protect \citeauthoryear {%
Blaschke%
, Fredriksson%
, Grigorian%
, {\"O}ztas%
\BCBL {}\ \BBA {} Sandin%
}{%
Blaschke%
\ \protect \BOthers {.}}{%
{\protect \APACyear {2005}}%
}]{%
Blaschke:2005uj}
\APACinsertmetastar {%
Blaschke:2005uj}%
\begin{APACrefauthors}%
Blaschke, D.%
, Fredriksson, S.%
, Grigorian, H.%
, {\"O}ztas, A.%
\BCBL {}\ \BBA {} Sandin, F.%
\end{APACrefauthors}%
\unskip\
\newblock
\APACrefYearMonthDay{2005}{}{},
\newblock
\unskip
\newblock
\APACjournalVolNumPages{Phys. Rev. D}{72}{}{065020}.
\PrintBackRefs{\CurrentBib}

\bibitem [\protect \citeauthoryear {%
Capano%
\ \protect \BOthers {.}}{%
Capano%
\ \protect \BOthers {.}}{%
{\protect \APACyear {2020}}%
}]{%
Capano:2019eae}
\APACinsertmetastar {%
Capano:2019eae}%
\begin{APACrefauthors}%
Capano, C\BPBI D.%
, Tews, I.%
, Brown, S\BPBI M.%
\ et al.\end{APACrefauthors}%
\unskip\
\newblock
\APACrefYearMonthDay{2020}{}{},
\newblock
\unskip
\newblock
\APACjournalVolNumPages{Nature Astron.}{4 (6)}{}{625}.
\PrintBackRefs{\CurrentBib}

\bibitem [\protect \citeauthoryear {%
Chatterjee%
\ \BBA {} Vida\~na%
}{%
Chatterjee%
\ \BBA {} Vida\~na%
}{%
{\protect \APACyear {2016}}%
}]{%
Chatterjee:2015pua}
\APACinsertmetastar {%
Chatterjee:2015pua}%
\begin{APACrefauthors}%
Chatterjee, D.%
\BCBT {}\ \BBA {} Vida\~na, I.%
\end{APACrefauthors}%
\unskip\
\newblock
\APACrefYearMonthDay{2016}{}{},
\newblock
\unskip
\newblock
\APACjournalVolNumPages{Eur. Phys. J. A}{52}{2}{29}.
\newblock
\begin{APACrefDOI} \doi{10.1140/epja/i2016-16029-x} \end{APACrefDOI}
\PrintBackRefs{\CurrentBib}

\bibitem [\protect \citeauthoryear {%
Cierniak%
\ \BBA {} Blaschke%
}{%
Cierniak%
\ \BBA {} Blaschke%
}{%
{\protect \APACyear {2020}}%
}]{%
Cierniak:2020eyh}
\APACinsertmetastar {%
Cierniak:2020eyh}%
\begin{APACrefauthors}%
Cierniak, M.%
\BCBT {}\ \BBA {} Blaschke, D.%
\end{APACrefauthors}%
\unskip\
\newblock
\APACrefYearMonthDay{2020}{}{},
\newblock
\unskip
\newblock
\APACjournalVolNumPages{Eur. Phys. J. ST}{229 (22-23)}{}{3663}.
\PrintBackRefs{\CurrentBib}

\bibitem [\protect \citeauthoryear {%
Cromartie%
\ \protect \BOthers {.}}{%
Cromartie%
\ \protect \BOthers {.}}{%
{\protect \APACyear {2020}}%
}]{%
Cromartie:2019kug}
\APACinsertmetastar {%
Cromartie:2019kug}%
\begin{APACrefauthors}%
Cromartie, H\BPBI T.%
, Fonseca, E.%
, Ransom, S\BPBI M.%
\ et al.\end{APACrefauthors}%
\unskip\
\newblock
\APACrefYearMonthDay{2020}{}{},
\newblock
\unskip
\newblock
\APACjournalVolNumPages{Nature Astron}{4 (1)}{}{72}.
\PrintBackRefs{\CurrentBib}

\bibitem [\protect \citeauthoryear {%
Drago%
\ \BBA {} Pagliara%
}{%
Drago%
\ \BBA {} Pagliara%
}{%
{\protect \APACyear {2020}}%
}]{%
Drago:2020gqn}
\APACinsertmetastar {%
Drago:2020gqn}%
\begin{APACrefauthors}%
Drago, A.%
\BCBT {}\ \BBA {} Pagliara, G.%
\end{APACrefauthors}%
\unskip\
\newblock
\APACrefYearMonthDay{2020}{}{},
\newblock
\unskip
\newblock
\APACjournalVolNumPages{Phys. Rev. D}{102}{6}{063003}.
\newblock
\begin{APACrefDOI} \doi{10.1103/PhysRevD.102.063003} \end{APACrefDOI}
\PrintBackRefs{\CurrentBib}

\bibitem [\protect \citeauthoryear {%
Fonseca%
\ \protect \BOthers {.}}{%
Fonseca%
\ \protect \BOthers {.}}{%
{\protect \APACyear {2021}}%
}]{%
Fonseca:2021wxt}
\APACinsertmetastar {%
Fonseca:2021wxt}%
\begin{APACrefauthors}%
Fonseca, E.%
\BCBT {}\ \BOthersPeriod {.}
\end{APACrefauthors}%
\unskip\
\newblock
\APACrefYearMonthDay{2021}{}{},
\newblock
\APACrefbtitle {{Refined Mass and Geometric Measurements of the High-Mass PSR
  J0740+6620}.} {{Refined Mass and Geometric Measurements of the High-Mass PSR
  J0740+6620}.}
\newblock
\APACrefnote{arXiv:2104.00880 [astro-ph.HE]}
\PrintBackRefs{\CurrentBib}

\bibitem [\protect \citeauthoryear {%
Glendenning%
}{%
Glendenning%
}{%
{\protect \APACyear {1992}}%
}]{%
Glendenning:1992vb}
\APACinsertmetastar {%
Glendenning:1992vb}%
\begin{APACrefauthors}%
Glendenning, N\BPBI K.%
\end{APACrefauthors}%
\unskip\
\newblock
\APACrefYearMonthDay{1992}{}{},
\newblock
\unskip
\newblock
\APACjournalVolNumPages{Phys. Rev. D}{46}{}{1274--1287}.
\newblock
\begin{APACrefDOI} \doi{10.1103/PhysRevD.46.1274} \end{APACrefDOI}
\PrintBackRefs{\CurrentBib}

\bibitem [\protect \citeauthoryear {%
Hinderer%
, Lackey%
, Lang%
\BCBL {}\ \BBA {} Read%
}{%
Hinderer%
\ \protect \BOthers {.}}{%
{\protect \APACyear {2010}}%
}]{%
Hinderer:2009ca}
\APACinsertmetastar {%
Hinderer:2009ca}%
\begin{APACrefauthors}%
Hinderer, T.%
, Lackey, B\BPBI D.%
, Lang, R\BPBI N.%
\BCBL {}\ \BBA {} Read, J\BPBI S.%
\end{APACrefauthors}%
\unskip\
\newblock
\APACrefYearMonthDay{2010}{}{},
\newblock
\unskip
\newblock
\APACjournalVolNumPages{Phys. Rev. D}{81}{}{123016}.
\newblock
\begin{APACrefDOI} \doi{10.1103/PhysRevD.81.123016} \end{APACrefDOI}
\PrintBackRefs{\CurrentBib}

\bibitem [\protect \citeauthoryear {%
Kl{\"a}hn%
, Lastowiecki%
\BCBL {}\ \BBA {} Blaschke%
}{%
Kl{\"a}hn%
\ \protect \BOthers {.}}{%
{\protect \APACyear {2013}}%
}]{%
Klahn:2013kga}
\APACinsertmetastar {%
Klahn:2013kga}%
\begin{APACrefauthors}%
Kl{\"a}hn, T.%
, Lastowiecki, R.%
\BCBL {}\ \BBA {} Blaschke, D.%
\end{APACrefauthors}%
\unskip\
\newblock
\APACrefYearMonthDay{2013}{}{},
\newblock
\unskip
\newblock
\APACjournalVolNumPages{Phys. Rev. D}{88 (8)}{}{085001}.
\PrintBackRefs{\CurrentBib}

\bibitem [\protect \citeauthoryear {%
Marczenko%
}{%
Marczenko%
}{%
{\protect \APACyear {2020}}%
}]{%
Marczenko:2020wlc}
\APACinsertmetastar {%
Marczenko:2020wlc}%
\begin{APACrefauthors}%
Marczenko, M.%
\end{APACrefauthors}%
\unskip\
\newblock
\APACrefYearMonthDay{2020}{}{},
\newblock
\unskip
\newblock
\APACjournalVolNumPages{Eur. Phys. J. ST}{229}{22-23}{3651--3661}.
\newblock
\begin{APACrefDOI} \doi{10.1140/epjst/e2020-000093-3} \end{APACrefDOI}
\PrintBackRefs{\CurrentBib}

\bibitem [\protect \citeauthoryear {%
Maslov%
\ \protect \BOthers {.}}{%
Maslov%
\ \protect \BOthers {.}}{%
{\protect \APACyear {2019}}%
}]{%
Maslov:2018ghi}
\APACinsertmetastar {%
Maslov:2018ghi}%
\begin{APACrefauthors}%
Maslov, K.%
, Yasutake, N.%
, Ayriyan, A.%
\ et al.\end{APACrefauthors}%
\unskip\
\newblock
\APACrefYearMonthDay{2019}{}{},
\newblock
\unskip
\newblock
\APACjournalVolNumPages{Phys. Rev. C}{100}{2}{025802}.
\newblock
\begin{APACrefDOI} \doi{10.1103/PhysRevC.100.025802} \end{APACrefDOI}
\PrintBackRefs{\CurrentBib}

\bibitem [\protect \citeauthoryear {%
Masuda%
, Hatsuda%
\BCBL {}\ \BBA {} Takatsuka%
}{%
Masuda%
\ \protect \BOthers {.}}{%
{\protect \APACyear {2013}}%
}]{%
Masuda:2012ed}
\APACinsertmetastar {%
Masuda:2012ed}%
\begin{APACrefauthors}%
Masuda, K.%
, Hatsuda, T.%
\BCBL {}\ \BBA {} Takatsuka, T.%
\end{APACrefauthors}%
\unskip\
\newblock
\APACrefYearMonthDay{2013}{}{},
\newblock
\unskip
\newblock
\APACjournalVolNumPages{PTEP}{2013}{7}{073D01}.
\newblock
\begin{APACrefDOI} \doi{10.1093/ptep/ptt045} \end{APACrefDOI}
\PrintBackRefs{\CurrentBib}

\bibitem [\protect \citeauthoryear {%
Miller%
\ \protect \BOthers {.}}{%
Miller%
\ \protect \BOthers {.}}{%
{\protect \APACyear {2019}}%
}]{%
Miller:2019cac}
\APACinsertmetastar {%
Miller:2019cac}%
\begin{APACrefauthors}%
Miller, M\BPBI C.%
, Lamb, F\BPBI K.%
, Dittmann, A\BPBI J.%
\ et al.\end{APACrefauthors}%
\unskip\
\newblock
\APACrefYearMonthDay{2019}{}{},
\newblock
\unskip
\newblock
\APACjournalVolNumPages{Astrophys. J. Lett.}{887 (1)}{}{L24}.
\PrintBackRefs{\CurrentBib}

\bibitem [\protect \citeauthoryear {%
Miller%
\ \protect \BOthers {.}}{%
Miller%
\ \protect \BOthers {.}}{%
{\protect \APACyear {2021}}%
}]{%
Miller:2021qha}
\APACinsertmetastar {%
Miller:2021qha}%
\begin{APACrefauthors}%
Miller, M\BPBI C.%
\BCBT {}\ \BOthersPeriod {.}
\end{APACrefauthors}%
\unskip\
\newblock
\APACrefYearMonthDay{2021}{}{},
\newblock
\APACrefbtitle {{The Radius of PSR J0740+6620 from NICER and XMM-Newton Data}.}
  {{The Radius of PSR J0740+6620 from NICER and XMM-Newton Data}.}
\newblock
\APACrefnote{arXiv:2105.06979 [astro-ph.HE]}
\PrintBackRefs{\CurrentBib}

\bibitem [\protect \citeauthoryear {%
Pang%
\ \protect \BOthers {.}}{%
Pang%
\ \protect \BOthers {.}}{%
{\protect \APACyear {2021}}%
}]{%
Pang:2021jta}
\APACinsertmetastar {%
Pang:2021jta}%
\begin{APACrefauthors}%
Pang, P\BPBI T\BPBI H.%
, Tews, I.%
, Coughlin, M\BPBI W.%
, Bulla, M.%
, Van Den~Broeck, C.%
\BCBL {}\ \BBA {} Dietrich, T.%
\end{APACrefauthors}%
\unskip\
\newblock
\APACrefYearMonthDay{2021}{}{},
\newblock
\APACrefbtitle {{Nuclear-Physics Multi-Messenger Astrophysics Constraints on
  the Neutron-Star Equation of State: Adding NICER's PSR J0740+6620
  Measurement}.} {{Nuclear-Physics Multi-Messenger Astrophysics Constraints on
  the Neutron-Star Equation of State: Adding NICER's PSR J0740+6620
  Measurement}.}
\newblock
\APACrefnote{arXiv:2105.08688 [astro-ph.HE]}
\PrintBackRefs{\CurrentBib}

\bibitem [\protect \citeauthoryear {%
Riley%
\ \protect \BOthers {.}}{%
Riley%
\ \protect \BOthers {.}}{%
{\protect \APACyear {2019}}%
}]{%
Riley:2019yda}
\APACinsertmetastar {%
Riley:2019yda}%
\begin{APACrefauthors}%
Riley, T\BPBI E.%
\BCBT {}\ \BOthersPeriod {.}
\end{APACrefauthors}%
\unskip\
\newblock
\APACrefYearMonthDay{2019}{}{},
\newblock
\unskip
\newblock
\APACjournalVolNumPages{Astrophys. J. Lett.}{887}{1}{L21}.
\newblock
\begin{APACrefDOI} \doi{10.3847/2041-8213/ab481c} \end{APACrefDOI}
\PrintBackRefs{\CurrentBib}

\bibitem [\protect \citeauthoryear {%
Riley%
\ \protect \BOthers {.}}{%
Riley%
\ \protect \BOthers {.}}{%
{\protect \APACyear {2021}}%
}]{%
Riley:2021pdl}
\APACinsertmetastar {%
Riley:2021pdl}%
\begin{APACrefauthors}%
Riley, T\BPBI E.%
\BCBT {}\ \BOthersPeriod {.}
\end{APACrefauthors}%
\unskip\
\newblock
\APACrefYearMonthDay{2021}{}{},
\newblock
\APACrefbtitle {{A NICER View of the Massive Pulsar PSR J0740+6620 Informed by
  Radio Timing and XMM-Newton Spectroscopy}.} {{A NICER View of the Massive
  Pulsar PSR J0740+6620 Informed by Radio Timing and XMM-Newton Spectroscopy}.}
\newblock
\APACrefnote{arXiv:2105.06980 [astro-ph.HE]}
\PrintBackRefs{\CurrentBib}

\bibitem [\protect \citeauthoryear {%
Shahrbaf%
, Blaschke%
\BCBL {}\ \BBA {} Khanmohamadi%
}{%
Shahrbaf%
\ \protect \BOthers {.}}{%
{\protect \APACyear {2020}}%
}]{%
Shahrbaf:2020uau}
\APACinsertmetastar {%
Shahrbaf:2020uau}%
\begin{APACrefauthors}%
Shahrbaf, M.%
, Blaschke, D.%
\BCBL {}\ \BBA {} Khanmohamadi, S.%
\end{APACrefauthors}%
\unskip\
\newblock
\APACrefYearMonthDay{2020}{}{},
\newblock
\unskip
\newblock
\APACjournalVolNumPages{J. Phys. G}{47}{11}{115201}.
\newblock
\begin{APACrefDOI} \doi{10.1088/1361-6471/abaa9a} \end{APACrefDOI}
\PrintBackRefs{\CurrentBib}

\bibitem [\protect \citeauthoryear {%
Somasundaram%
\ \BBA {} Margueron%
}{%
Somasundaram%
\ \BBA {} Margueron%
}{%
{\protect \APACyear {2021}}%
}]{%
Somasundaram:2021ljr}
\APACinsertmetastar {%
Somasundaram:2021ljr}%
\begin{APACrefauthors}%
Somasundaram, R.%
\BCBT {}\ \BBA {} Margueron, J.%
\end{APACrefauthors}%
\unskip\
\newblock
\APACrefYearMonthDay{2021}{}{},
\newblock
\APACrefbtitle {{Impact of massive neutron star radii on the nature of phase
  transitions in dense matter}.} {{Impact of massive neutron star radii on the
  nature of phase transitions in dense matter}.}
\newblock
\APACrefnote{arXiv:2104.13612 [astro-ph.HE]}
\PrintBackRefs{\CurrentBib}

\bibitem [\protect \citeauthoryear {%
Tan%
, Dore%
, Dexheimer%
, Noronha-Hostler%
\BCBL {}\ \BBA {} Yunes%
}{%
Tan%
\ \protect \BOthers {.}}{%
{\protect \APACyear {2021}}%
}]{%
Tan:2021ahl}
\APACinsertmetastar {%
Tan:2021ahl}%
\begin{APACrefauthors}%
Tan, H.%
, Dore, T.%
, Dexheimer, V.%
, Noronha-Hostler, J.%
\BCBL {}\ \BBA {} Yunes, N.%
\end{APACrefauthors}%
\unskip\
\newblock
\APACrefYearMonthDay{2021}{}{},
\newblock
\APACrefbtitle {{Extreme Matter meets Extreme Gravity: Ultra-heavy neutron
  stars with crossovers and first-order phase transitions}.} {{Extreme Matter
  meets Extreme Gravity: Ultra-heavy neutron stars with crossovers and
  first-order phase transitions}.}
\newblock
\APACrefnote{arXiv:2106.03890 [astro-ph.HE]}
\PrintBackRefs{\CurrentBib}

\bibitem [\protect \citeauthoryear {%
Yamamoto%
, Furumoto%
, Yasutake%
\BCBL {}\ \BBA {} Rijken%
}{%
Yamamoto%
\ \protect \BOthers {.}}{%
{\protect \APACyear {2016}}%
}]{%
Yamamoto:2015lwa}
\APACinsertmetastar {%
Yamamoto:2015lwa}%
\begin{APACrefauthors}%
Yamamoto, Y.%
, Furumoto, T.%
, Yasutake, N.%
\BCBL {}\ \BBA {} Rijken, T\BPBI A.%
\end{APACrefauthors}%
\unskip\
\newblock
\APACrefYearMonthDay{2016}{}{},
\newblock
\unskip
\newblock
\APACjournalVolNumPages{Eur. Phys. J. A.}{52 (2)}{}{19}.
\PrintBackRefs{\CurrentBib}

\bibitem [\protect \citeauthoryear {%
Yamamoto%
\ \protect \BOthers {.}}{%
Yamamoto%
\ \protect \BOthers {.}}{%
{\protect \APACyear {2017}}%
}]{%
Yamamoto:2017wre}
\APACinsertmetastar {%
Yamamoto:2017wre}%
\begin{APACrefauthors}%
Yamamoto, Y.%
, Togashi, H.%
, Tamagawa, T.%
, Furumoto, T.%
, Yasutake, N.%
\BCBL {}\ \BBA {} Rijken, T\BPBI A.%
\end{APACrefauthors}%
\unskip\
\newblock
\APACrefYearMonthDay{2017}{}{},
\newblock
\unskip
\newblock
\APACjournalVolNumPages{Phys. Rev. C}{96 (6)}{}{065804}.
\PrintBackRefs{\CurrentBib}

\bibitem [\protect \citeauthoryear {%
Yudin%
, Razinkova%
, Nadyozhin%
\BCBL {}\ \BBA {} Dolgov%
}{%
Yudin%
\ \protect \BOthers {.}}{%
{\protect \APACyear {2014}}%
}]{%
Yudin:2014mla}
\APACinsertmetastar {%
Yudin:2014mla}%
\begin{APACrefauthors}%
Yudin, A.%
, Razinkova, T\BPBI L.%
, Nadyozhin, D\BPBI K.%
\BCBL {}\ \BBA {} Dolgov, A\BPBI D.%
\end{APACrefauthors}%
\unskip\
\newblock
\APACrefYearMonthDay{2014}{}{},
\newblock
\unskip
\newblock
\APACjournalVolNumPages{Astron. Lett.}{40}{}{201}.
\PrintBackRefs{\CurrentBib}

\bibitem [\protect \citeauthoryear {%
Zdunik%
\ \BBA {} Haensel%
}{%
Zdunik%
\ \BBA {} Haensel%
}{%
{\protect \APACyear {2013}}%
}]{%
Zdunik:2012dj}
\APACinsertmetastar {%
Zdunik:2012dj}%
\begin{APACrefauthors}%
Zdunik, J\BPBI L.%
\BCBT {}\ \BBA {} Haensel, P.%
\end{APACrefauthors}%
\unskip\
\newblock
\APACrefYearMonthDay{2013}{}{},
\newblock
\unskip
\newblock
\APACjournalVolNumPages{Astron. Astrophys.}{551}{}{A61}.
\PrintBackRefs{\CurrentBib}

\end{thebibliography}
\end{document}